\documentstyle[12pt,aas2pp4]{article}

\begin{document}
\input epsf

\begin{flushright}
PSU-PAL-97-1
\end{flushright}

\vskip 0.4cm

\title {The Energy Spectra and Relative Abundances of Electrons\\
and Positrons in the Galactic Cosmic Radiation}

\author{S.~W.~Barwick,\altaffilmark{1} J.~J.~Beatty,\altaffilmark{2}
C.~R.~Bower,\altaffilmark{3} C.~J.~Chaput,\altaffilmark{4}
S.~Coutu,\altaffilmark{2,4} G.~A.~de Nolfo,\altaffilmark{2}
M.~A.~DuVernois,\altaffilmark{2} D.~Ellithorpe,\altaffilmark{5}
D.~Ficenec,\altaffilmark{6} J.~Knapp,\altaffilmark{5,7}
D.~M.~Lowder,\altaffilmark{8} S.~McKee, \altaffilmark{4}
D.~M\"uller,\altaffilmark{5} J.~A.~Musser,\altaffilmark{3}
S.~L.~Nutter,\altaffilmark{2,9} E.~Schneider,\altaffilmark{1}
S.~P.~Swordy,\altaffilmark{5}
G.~Tarl\'e,\altaffilmark{4} A.~D.~Tomasch,\altaffilmark{4}
and E.~Torbet\altaffilmark{5}}

\altaffiltext{1} {Department of Physics, University of California at Irvine,
Irvine, CA 92717.}
\altaffiltext{2} {Departments of Physics and of Astronomy and Astrophysics, 
104 Davey Laboratory, Pennsylvania State University, University Park, 
PA 16802.}
\altaffiltext{3} {Department of Physics, Swain Hall West, Indiana University,
Bloomington, IN 47405.}
\altaffiltext{4} {Department of Physics, Randall Laboratory, 500 E. 
University, University of Michigan,
Ann Arbor, MI 48109-1120.}
\altaffiltext{5} {Enrico Fermi Institute and Department of Physics, 
933 E. 56$^{\rm th}$ St., University of Chicago,
Chicago, IL 60637.}
\altaffiltext{6} {Present address: Millenium Pharmaceuticals, Inc., 640
Memorial Drive, Cambridge, MA 02139.}
\altaffiltext{7} {Present address: Institut f\"ur Experimentelle Kernphysik,
Universit\"at Karlsruhe, Postfach 3640, D 76021 Karlsruhe, Germany.}
\altaffiltext{8} {Physics Department, 366 LeConte Hall, University of 
California at Berkeley, Berkeley, CA 94720.}
\altaffiltext{9} {Present address: Department of Physical Sciences, Eastern 
New Mexico University, Portales, NM 88130.}

\begin{abstract}
Observations of cosmic-ray electrons and positrons have been
made with a new balloon-borne detector, HEAT (the ``High-Energy
Antimatter Telescope''), first flown in 1994 May from Fort Sumner, NM.
We describe the instrumental approach and the data analysis procedures, and
we present results from this flight. The measurement has provided a
new determination of the individual energy spectra of electrons and
positrons from 5~GeV to about 50~GeV,
and of the combined ``all-electron'' intensity ${\rm (e^++e^-)}$ up to 
$\sim 100$~GeV. The single power-law spectral indices for electrons and 
positrons are $\alpha = 3.09 \pm 0.08$
and $3.3 \pm 0.2$, respectively. 
We find that a contribution from primary sources to the positron intensity
in this energy region, if it exists, must be quite small.

\end{abstract}


\section{Introduction}

Electrons and positrons are a relatively rare component of the cosmic
radiation, but are distinct from all other cosmic-ray particles because
of their 
low mass and the absence of hadronic interactions with the constituents of the
interstellar medium. Previous measurements of electrons ${\rm (e^+ +e^-)}$ up 
to roughly 1000~GeV (\cite{pri:the,nis:ele,gol:ele,tan:the}) have shown that 
their intensity amounts to about
1\% of the flux of protons around 10~GeV, but decreases more rapidly with
energy ($\propto E^{-3.1}$) than the proton spectrum ($\propto
E^{-2.7}$). Separate measurements of positrons and electrons have only been
possible at much lower energies and have indicated a ``positron fraction'' 
(${\rm e^+/(e^++e^-)}$) of a few percent in the region 1-10~GeV 
(\cite{fan:pos,buf:pos,gol:ts93,bar:cap}). Some measurements have indicated an
increase above about 10~GeV (\cite{agr:pos,mul:pos,go1:pos,go2:pos}).
These observations have led to a number of conclusions, but have also left some
key questions unanswered:

1) The predominance of negative electrons can only be explained if electrons
are accelerated by primary sources. The alternate production mechanism, as
secondary particles from interstellar nuclear interactions of hadronic cosmic
rays (mostly through the $\pi^\pm \rightarrow \mu^\pm \rightarrow 
{\rm e}^\pm $
decay), would yield negative and positive electrons in about equal proportions,
and thus can only account for a small portion of the total ${\rm e^\pm}$ flux.
It is worth pointing out that electrons are the only cosmic-ray component for
which an extragalactic contribution can be excluded with certainty: 
energy losses
through Compton interactions with the 2.7K background radiation preclude their
propagation through intergalactic distances.

2) The steepness of the observed energy spectrum of electrons is usually
explained as a consequence of radiative energy losses during propagation
through the interstellar medium: inverse Compton scattering with
photons and synchrotron radiation in the
interstellar magnetic fields. Assuming that electrons are accelerated with the
same energy spectrum at the source as nuclei ($\propto E^{-2.15}$,
\cite{mul:sou,swo:sou}) the observed spectral slope is roughly consistent with the
galactic containment time of $\sim 10^7$~years of nuclei at GeV energies.
However, there are several problems with this interpretation: First, the energy
spectrum of electrons at the source is not known a priori, nor is it known that
electrons and nuclei originate at the same acceleration sites. Second, it has
been pointed out (\cite{tan:the}) that the energy dependence ($\propto E^{-0.6}$)
of the containment time observed for nuclei may be difficult to reconcile with
the observed shape of the electron spectrum. Third, the ``leaky box'' assumption
inherent in these explanations should not be applied to electrons, as it 
requires an unreasonably high density of electron sources in the 
galactic disk (\cite{cow:ele}).

3) The small flux of positrons appears to be essentially consistent with a 
secondary origin.
If this is true, the production spectrum of positrons can be calculated from 
the known flux of primary nuclei (\cite{pro:pos}). Positrons would be produced
continuously throughout the galactic disk, and the leaky box model could more
easily be taken as an approximation for their propagation through the galaxy.
In any case, a direct comparison between the production spectrum and the
observed positron spectrum would permit a quantitative study of the propagation
mechanism. However, it must first be ascertained that indeed all positrons are
of secondary origin. The previously observed increase in the positron fraction
has led to much speculation about the appearance of primary positrons at high 
energy (e.g. \cite{har:pos,aha:pos,dog:pos,tyl:pos,tur:dir,tur:ind}). 
Recent results
(\cite{bar:prl,bar:apjl}) have not confirmed this increase, but have not
yet shown conclusively that the positron flux is entirely free of primary
contributions.

It is clear from these considerations that many of the open questions can only
be answered through measurements of electrons and positrons separately, and
over as large an energy range as possible. This was the motivation for the
construction of the High-Energy Antimatter Telescope (HEAT), an instrument that
utilizes a large superconducting magnet spectrometer to separate positive and
negative particles, and that incorporates powerful techniques to identify
electrons and positrons and to reject hadronic background. A first balloon
flight of HEAT was conducted in 1994. In the following, we shall describe this
measurement and the data analysis technique, and we shall present and discuss
the results.

\section{Flight}

\begin{figure}
\vspace{-0.5cm}
\hbox to \textwidth{\hss
\epsfxsize=0.4\textwidth
\epsfbox[190 310 430 530]{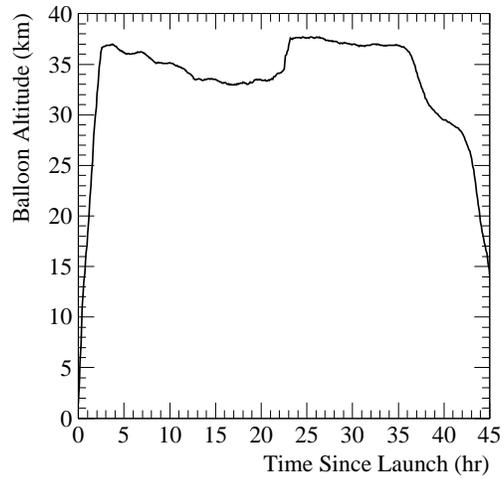}
\hss}
\caption{\it Altitude profile of the 1994 balloon flight
\label{fig1} }
\end{figure}

The first balloon flight of the HEAT e$^\pm$ instrument took place on
1994 May 3-5, from Fort Sumner, New Mexico, and data were collected at float 
altitude for about 29~hours.
The payload reached a maximum altitude of 36.5~km and
drooped to a minimum of 33~km at night, as illustrated in Figure~\ref{fig1}.
Over the course of the flight, the payload drifted between vertical
geomagnetic cutoff rigidities of 4~GV and 4.5~GV,
latitudes of $33.35^{\circ}$~N and $35.3^{\circ}$~N and longitudes of
$100.0^{\circ}$~W and $104.3^{\circ}$~W.
The instrument was recovered undamaged near Wellington, Texas.

\placefigure{fig1}

\section{Instrument Description and Performance}

\begin{figure}
\vspace{-0.5cm}
\hbox to \textwidth{\hss
\epsfxsize=0.4\textwidth
\epsfbox[110 150 540 655]{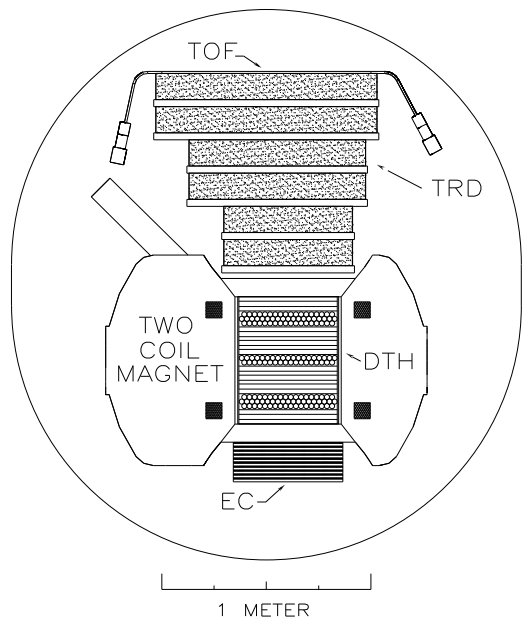}
\hss}
\caption{\it HEAT instrument schematic cross-section
\label{fig2} }
\end{figure}

The expected intensity of cosmic-ray po\-si\-trons is quite low, of the order of
$10^{-4}$ of the total cosmic-ray intensity at comparable energies. A
successful measurement then necessitates a detector with large sensitive area
to yield statistically significant data, and with very high discrimination
power against the overwhelming proton background. This is accomplished in the
HEAT instrument, shown in Figure~\ref{fig2}, through the combination of a 
superconducting magnet spectrometer (using a drift-tube hodoscope (DTH) 
tracking chamber) with particle identifiers employing a time-of-flight
system (TOF), a transition-radiation detector (TRD), and an electromagnetic 
ca\-lo\-ri\-me\-ter (EC). A thorough description of the instrument appears
elsewhere (\cite{heat:nim}), and the following just gives a brief summary of 
the apparatus and its performance.

\placefigure{fig2}

\subsection{Time-of-Flight System}

The TOF measures the velocity of the particle, 
distinguishes downward from upward-going (albedo) particles, and measures the 
magnitude of the particle's charge. The rejection of upward-going particles is 
necessary as these would otherwise appear as particles of the opposite charge
in the magnet spectrometer. Through the measurement of the magnitude of the 
charge, singly-charged particles are identified and discriminated from
helium and heavier nuclei.

The TOF consists of four scintillator slabs on top of the instrument, 
each with an active area of 100~cm $\times$ 25~cm, and of
the top three scintillation counters of the EC (described below).
Throughout the instrument, photomultiplier tubes (PMT) are used that are
resistant to the magnetic fringe fields.

\begin{figure}
\vspace{-0.5cm}
\hbox to \textwidth{\hss
\epsfxsize=0.8\textwidth
\epsfbox[65 305 545 540]{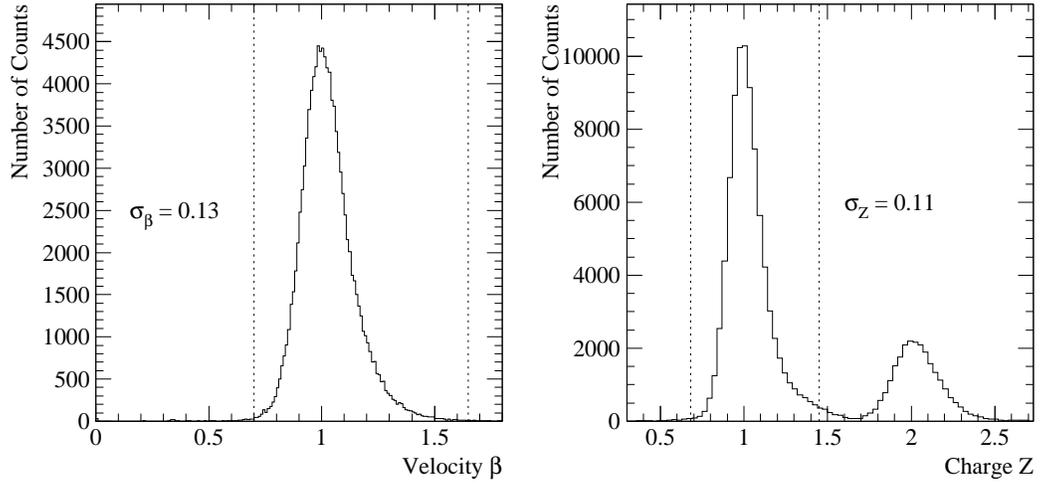}
\hss}
\caption{\it TOF $\beta$ and charge distributions
\label{fig3} }
\end{figure}

The PMT signals are both charge- (ADC) and time- (TDC) digitized, and the
particle's charge is determined from the ADC values measured with the top
scintillator slabs, each of which has a PMT on either end.
Figure~\ref{fig3} shows the charge distribution obtained for flight data, 
with the criteria for selecting singly-charged particles
shown as dashed lines (and summarized in Table~\ref{all_cuts}).
The charge resolution obtained is 0.11 charge units, which is sufficient to 
provide good 
rejection of He events. The TDC values, after corrections for the path length 
through the instrument, give the particle's velocity, with a resolution of 
0.13~c (where c is the speed of light). Figure~\ref{fig3} shows the velocity 
distribution obtained for flight data, and indicates the selection criteria
used for the analysis. 

\placefigure{fig3}

\subsection{Transition-Radiation Detector}

The transition-radiation detector is used to
distinguish electrons and positrons from hadrons.
It is comprised of 6 modules, each consisting of a polyethylene-fiber
radiator and a multiwire proportional chamber (MWPC). The MWPCs contain
a gas mixture of xe\-non:\-me\-tha\-ne (70:30) and produce signals on both the
anode wires and on cathode strips. The total charge deposited in the chamber 
is read from the cathode strips and pulse-height analyzed (``PHA'' analysis),
while the current signal is read from the anode wires in 25~nsec-wide time 
slices, which are digitized into three different threshold levels
(``time-slice'' analysis).

Charged particles of high Lorentz factor ($\gamma > 10^3$) produce 
transition-radiation (TR) x-rays (5--30~keV) in the radiator, which are 
subsequently detected by the MWPC. This x-ray signal is superimposed on the 
signal due to ionization energy loss of the parent particle. In the energy 
range of 5--50~GeV, electrons produce a saturated TR signal, while protons 
and pions produce none, thereby permitting a discrimination between 
these spe\-cies.

\subsubsection{PHA Analysis}

\begin{figure}
\vspace{-0.4cm}
\hbox to \textwidth{\hss
\epsfxsize=0.38\textwidth
\epsfbox[190 200 430 650]{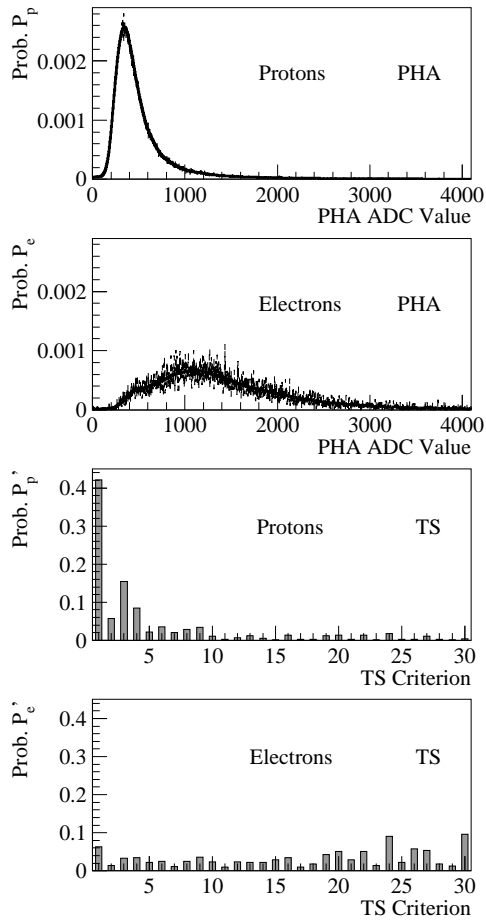}
\hss}
\caption{\it PHA and Time-Slice probability distributions for
protons and electrons
\label{fig4} }
\end{figure}

A maximum-likelihood technique is used to analyze the PHA signals
(\cite{che:trd}). To this end, first the probability distributions for the
pulse heights in the MWPC's must be determined,
using clean populations of electrons and protons. The electron sample is 
obtained by selecting events with negative rigidity in the magnet 
spectrometer (described below), and a shower profile in the EC consistent with 
an electromagnetic shower (also described below). The proton sample is obtained by 
selecting events with positive rigidity and which have shower profiles 
inconsistent with an electromagnetic shower. The validity of this procedure
has been verified with accelerator calibrations at Fermilab. Gain 
differences between the TRD chambers are 
corrected for, as are temporal variations (30\%) in 
the chamber gains caused by temperature and pressure
changes inside the instrument gondola, spatial non-uniformities (20\%), and
the relativistic rise of the proton ionization signal.
The upper two panels of Figure~\ref{fig4} show the PHA probability
distributions P$_{\rm e}$ for electrons and P$_{\rm p}$ for protons. 

\placefigure{fig4}

\begin{figure}
\vspace{-0.5cm}
\hbox to \textwidth{\hss
\epsfxsize=0.4\textwidth
\epsfbox[180 310 430 540]{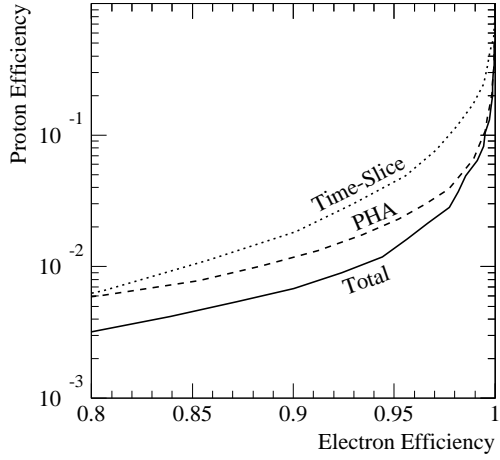}
\hss}
\caption{\it TRD proton and electron efficiencies
\label{fig5} }
\end{figure}

For the subsequent data analysis, each e\-vent, characterized by the MWPC
pulse hei\-ghts $x_i$ ($i=$1,...,6), is compared with the probability
distributions, and a likelihood ratio $L_{\mbox{\scriptsize \it PHA}}$ is
formed:
\[ L_{\mbox{\scriptsize \it PHA}} = \prod_{i=1}^{6}
\frac{P_e(x_i)}{P_p(x_i)}. \]
Figure~\ref{fig5} shows as a dashed line labeled ``PHA'' the electron and 
proton efficiencies obtained by applying a selection on the PHA likelihood 
ratio. 

\placefigure{fig5}

\subsubsection{Time-Slice Analysis}

The time-slice technique makes use of the fact that an x-ray photon absorbed
in the MWPC generates a highly localized ``ionization cluster'' in the
chamber, which then drifts to the anode wires and produces a ``spike'' in the
time structure of the anode signal. In the analysis, again, a likelihood ratio 
$L'_{\mbox{\scriptsize \it TS}}$ is constructed, using the appropriate 
single-chamber probability distributions $P'_e$ and $P'_p$.
These distributions are based on the number of time slices above each 
threshold level, and the positions, heights, and number of clusters in the
time-slice distribution, after the gain corrections described above are applied.
The lower two panels of Figure~\ref{fig4} show the time-slice probability
distributions obtained for electrons and protons. Figure~\ref{fig5} shows as
a dotted line labeled ``Time-Slice'' the electron and proton efficiencies 
obtained by applying a selection on the time-slice likelihood ratio.

\begin{figure}
\vspace{-0.5cm}
\hbox to \textwidth{\hss
\epsfxsize=0.4\textwidth
\epsfbox[180 310 430 530]{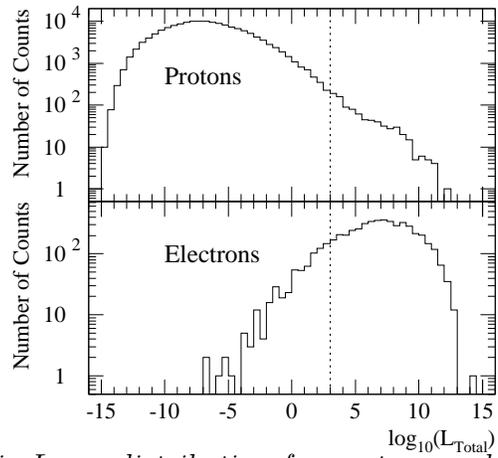}
\hss}
\caption{\it TRD likelihood ratio 
$L_{\mbox{\scriptsize \it Total}}$ distribution for protons and electrons
\label{fig6} }
\end{figure}

The likelihood ratios  $L_{\mbox{\scriptsize \it PHA}}$ and 
$L'_{\mbox{\scriptsize \it TS}}$, are combined to form a total 
ratio: $L_{\mbox{\scriptsize \it Total}} = L_{\mbox{\scriptsize 
\it PHA}} \cdot L'_{\mbox{\scriptsize \it TS}}$.
Figure~\ref{fig6} shows the distribution of $L_{\mbox{\scriptsize 
\it Total}}$ together with the selection criterion used to identify
electrons ($L_{\mbox{\scriptsize \it Total}} > 10^3$, or dashed line in the
figure). With this criterion, electrons are retained
with an efficiency $\epsilon_e = 88\%$ and protons with an 
efficiency $\epsilon_p = 0.59\%$, corresponding to a proton rejection power
of 170. This is illustrated in Figure~\ref{fig5} as a solid line labeled 
``Total.''

\placefigure{fig6}

\subsection{Magnet Spectrometer \label{DTH}}

The magnet spectrometer is used to measure the rigidity $R = {pc \over Ze}$ 
and charge sign of a traversing charged particle, and consists of a two-coil
warm-bore superconducting magnet and a precision tracking detector utilizing
drift tubes. The magnet produces an approximately uniform field of central
value 1~T. The fiducial volume
within the magnet bore measures 50~cm $\times$ 50~cm $\times$ 61~cm.
The tracking system consists of 479 drift tubes of 2.5~cm diameter,
arranged in 26 rows; 
18 rows (of which 17 were operational during the flight) contain tubes 
parallel to the magnet axis, defining the bending view, and 8 contain
tubes perpendicular to the magnet axis, defining the non-bending view. The 
drift gas mixture used is CO$_{2}$:hexane (96:4).

\begin{figure}
\vspace{-0.5cm}
\hbox to \textwidth{\hss
\epsfxsize=0.4\textwidth
\epsfbox[180 310 430 530]{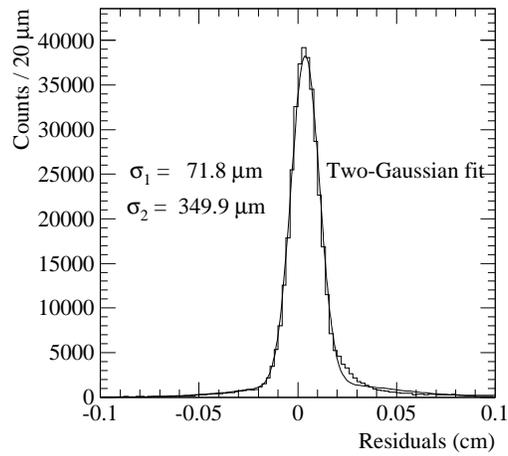}
\hss}
\caption{\it Distribution of DTH residuals for bending-view tubes
\label{fig7} }
\end{figure}

\begin{figure}
\vspace{-0.5cm}
\hbox to \textwidth{\hss
\epsfxsize=0.4\textwidth
\epsfbox[180 310 430 530]{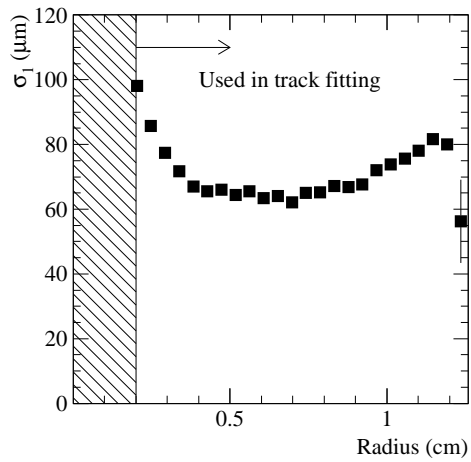}
\hss}
\caption{\it DTH residuals as a function of tube radius
for bending-view tubes
\label{fig8} }
\end{figure}

The DTH determines the particle trajectory by measuring the drift times of
the ionization tracks in each tube hit, from which, using a ``time-to-space''
function, the impact parameters (distance of closest approach to wire) are
found. The trajectory of the particle in three dimensions is reconstructed
from the impact parameters using a modified version of the CERN program 
libraries' MOMENTM algorithm (\cite{win:mom}). Figures~\ref{fig7}
and~\ref{fig8}
show the residuals from the tracks reconstructed for flight data, as a
distribution and as a function of radius, respectively. A signal is not
used in the track fit if it lies more than one tube diameter away
from the reconstructed trajectory,
or if it has a residual that is worse than 6 times the average residual
for the track. Also removed are signals corresponding to an impact 
parameter smaller than
2~mm, a region where the tracking capabilities of the drift tubes are poor
due to the nature of the electron-ion pair statistics and the increased
drift speed near the wire.
The single-tube resolution achieved is about 70~$\mu$m.

\placefigure{fig7}

\placefigure{fig8}

\begin{figure}
\vspace{-0.5cm}
\hbox to \textwidth{\hss
\epsfxsize=0.4\textwidth
\epsfbox[180 310 430 530]{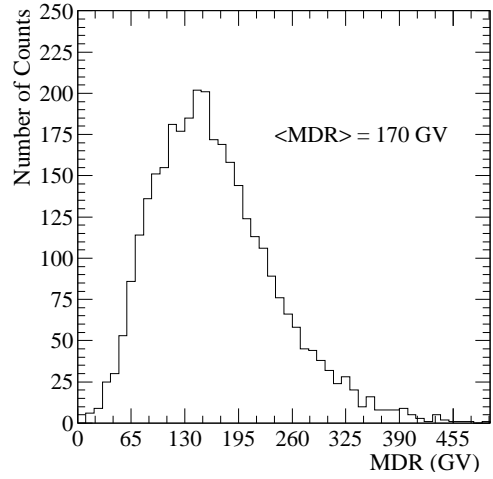}
\hss}
\caption{\it MDR distribution for electron events
\label{fig9} }
\end{figure}

The performance of the magnet spectrometer is characterized by the maximum
detectable rigidity (MDR). The MDR is defined as that rigidity where
the RMS error in the sagitta measurement is equal to
the sagitta of the track, and is computed on an event-by-event
basis.  The MDR distribution for electron events, given in Figure~\ref{fig9}, 
shows that a mean MDR of 170~GV is achieved.
The MDR yields the relative RMS error in the rigidity:
$\sigma_R/R = R/MDR$. Thus, an MDR of 170~GV provides for a 
rigidity determination with an at least $3\sigma$ accuracy up to $\sim$~60~GV.
In the analysis of the flight data, the selections $R/MDR < 0.25$ and 
$MDR > 60$ are imposed to ensure meaningful results.

\placefigure{fig9}

\subsection{Electromagnetic Calorimeter}

The electromagnetic calorimeter measures the energy of electrons and 
discriminates a\-gainst hadrons. The EC consists of ten lead plates, each
0.9~radiation length thick and 50~cm $\times$ 50~cm in area, and each followed
by a plastic scintillator. 

Electrons deposit most of their energy in the EC, with well-understood shower
profiles that provide a measure of their energy. Protons and other hadrons,
on the other hand, rarely interact in the EC, as it represents only about 0.3 
proton interaction lengths, and if a nuclear interaction occurs, the 
longitudinal shower
profile is usually quite different from that of an electron. Therefore, the EC 
also provides powerful discrimination against protons.

\begin{figure}
\vspace{-0.5cm}
\hbox to \textwidth{\hss
\epsfxsize=0.4\textwidth
\epsfbox[180 310 430 530]{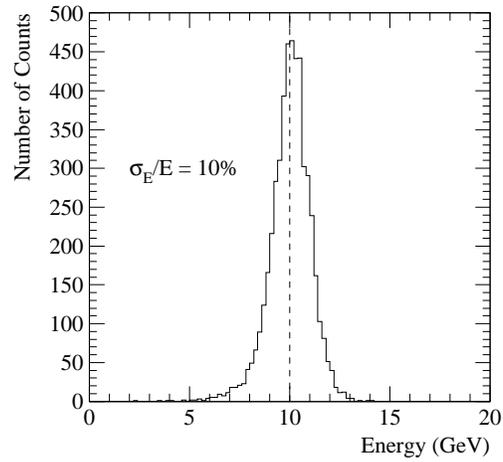}
\hss}
\caption{\it EC energy distribution for 10~GeV MC events
\label{fig10} }
\end{figure}

In order to reconstruct the electron primary energy from the shower profile,
Monte Carlo simulations have been carried out using the CERN program 
li\-bra\-ries' GEANT pack\-a\-ge (\cite{bru:gea}). The results of these
simulations have been verified through accelerator calibrations 
(\cite{tor:cal}). A covariance analysis is applied to the Monte Carlo events, 
providing the cross-correlations between the signals obtained in each of the
layer pairs as a function of energy.
Figure~\ref{fig10} shows, for 10~GeV simulated electrons, the
distribution of energies reconstructed by this covariance analysis. 
The correlation matrices generated by the covariance analysis of 
Monte Carlo events are then applied to flight data to determine the particle 
energies. The fractional energy resolution achieved is about 10\% for the
energy range of interest.

\placefigure{fig10}

\begin{figure}
\vspace{-0.5cm}
\hbox to \textwidth{\hss
\epsfxsize=0.4\textwidth
\epsfbox[180 310 430 530]{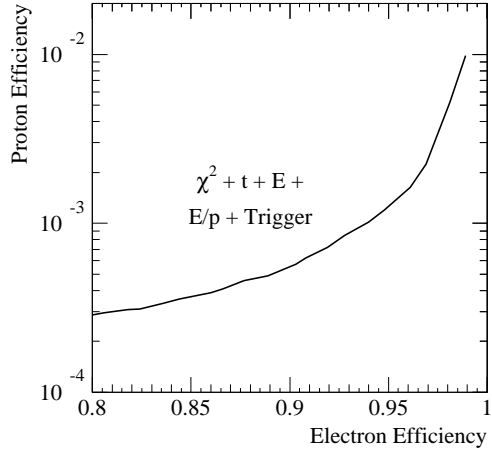}
\hss}
\caption{\it EC proton and electron efficiencies
\label{fig11} }
\end{figure}

During the balloon flight, an event trigger is employed (see below) that
requires a minimum energy deposition in the EC, and thereby removes most
non-interacting protons. In order to discriminate against the remaining proton 
events with the EC, a $\chi^2$ measure of the agreement between the observed 
and expected shower profile, generated by the covariance analysis, is 
determined for each event. In addition, the shower start depth $t$ is derived
from a separate profile analysis of the shower, and events with $t > 0.89$
radiation length are rejected. Finally, agreement between the energy $E$ 
measured by the EC, and the momentum $p$ measured by the magnet spectrometer is
demanded. Figure~\ref{fig11} shows the electron and proton efficiencies 
obtained for flight data by imposing all of these requirements. The selection
criteria chosen for the analysis are summarized in Table~\ref{all_cuts}, and
yield for the combination of shower counter and magnet spectrometer a 
proton rejection power of 460 at an electron efficiency of 97\%. Combined with 
that of the TRD (see above), the total rejection power of the instrument 
against protons is nearly $10^5$. 

\placefigure{fig11}

\subsection{Instrument Trigger}

The instrument trigger requires a through-going particle which deposits 
a signal larger than that of a 0.5~GeV electron in the EC. This excludes
non-interacting protons. In addition, a minimum number of hits in the DTH is 
required. A ``fast'' trigger is formed by the coincidence 
of signals in the top and bottom TOF scintillators, and of a signal above 
threshold in the sum of the lower seven EC layers. A ``slow'' confirming 
trigger is required within 1.5 $\mu$sec thereafter, based on the occurrence 
of a three-fold majority of hits in the first, fifth, and last two rows of
tubes in the bending plane. In order to obtain a sample of non-interacting 
protons, the EC sum threshold requirement is removed for a small 
percentage (2\%) of the events.

\section{Electron Selection}

\subsection{Event Filtering}

The initial data analysis stage involves examining the DTH hits for clean
single-particle events. 
A rough estimate of the trajectory is made by performing a quadratic fit to 
the impact parameters in both the bending and non-bending views. Signals 
from tubes well 
outside this fit are discarded. Events in which tracks cannot be 
identified in the DTH through this procedure are rejected. Most events 
eliminated at this stage result from interactions of particles that penetrate
the instrument from the sides.

Tracks which pass the initial DTH filter are then analyzed using 
the modified MOMENTM algorithm, which determines the 
particle's rigidity from the measured track points. A $\chi^2$ parameter, 
generated by comparing the measured track points with the fit, is used to 
reject events with large tracking errors. In addition, events are rejected 
based on the average residuals associated with the track, as well as on the 
number of tubes retained in the final fit in the bending and non-bending 
views. Finally, the MDR requirements described in Section~\ref{DTH} are 
imposed. Table~\ref{all_cuts} summarizes all of the track cleanliness 
requirements.

\placetable{all_cuts}

\subsection{Template Fits}

Electron selection criteria are applied to events that satisfy the track 
cleanliness requirements. As discussed, these include selections on the
magnitude of the particle charge, the TRD total likelihood ratio, the 
EC covariance 
analysis $\chi^2$ parameter, the shower start depth, and reconstructed energy.
In order to maximize the statistics of the results, the electron selection 
criteria have been chosen relatively loosely, requiring that the number of 
background protons be less than 25\% of the number of accepted positrons. 
Table~\ref{all_cuts} summarizes the selections used.

\begin{figure}
\vspace{-0.5cm}
\hbox to \textwidth{\hss
\epsfxsize=0.4\textwidth
\epsfbox[180 170 430 450]{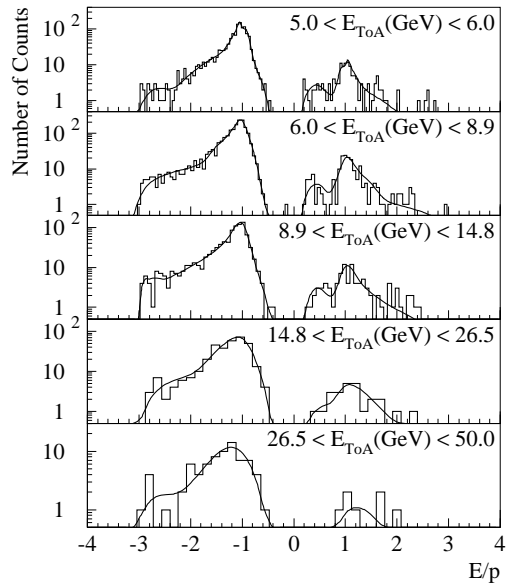}
\hss}
\caption{\it $E/p$ distributions for electrons and positrons
\label{fig12} }
\end{figure}

For events meeting the electron selection criteria, distributions of the ratio
of the measured
energy $E$ determined by the EC, and the momentum $p$ determined by the
magnet spectrometer, are generated. Figure~\ref{fig12} shows these $E/p$ 
distributions for electrons and positrons in five  
top-of-the-atmosphere (ToA) energy intervals. (The 
top-\-of-\-the-\-at\-mos\-phe\-re
energy $E_{ToA}$ is obtained from the measured energy $E$ by correcting 
for atmospheric brems\-strah\-lung losses according to the procedure described in
Section~\ref{atmoscor}.)
The differences in $E/p$ between positrons and re\-si\-dual background protons is 
evident, with the positron distribution peaking at +1, and that for protons 
peaking at about +0.5. The tails at high $\left| E/p \right|$ values in the 
electron or positron distributions are due to brems\-strah\-lung losses in the
material above the DTH, which result in a reduced rigidity in the spectrometer 
while the entire energy (of the e$^\pm$ plus bremsstrahlung photons) is
usually recorded in the EC.

\placefigure{fig12}

A fit to the $E/p$ distributions for electrons and positrons is performed 
to obtain the 
number of electrons and positrons, and to estimate and subtract the residual 
proton background. The $E/p$ distribution for negative electrons, which can be
taken as background free, is used to obtain a template of the expected $E/p$ 
distribution for e$^\pm$. The $E/p$ distribution for background protons
is obtained by inverting 
the TRD selection to select interacting protons (see Table~\ref{all_cuts}). 
Both template distributions are smoothed, and used to fit the $E/p$ 
distribution for positrons, shown as the solid curves in Figure~\ref{fig12}. 
A Bayesian treatment is then used to estimate the electron, positron and
proton counts,
with prior probability distributions assumed to be flat for the electron, 
positron and proton background counts, and with conditional distributions taken 
from the template fits. The electron and positron counts resulting from this 
procedure are listed in Table~\ref{elec_numb}. The uncertainties given in the 
table are 16 and 84\% Bayesian limits.
For the energy range 50--100~GeV, the ability of the magnetic spectrometer to
determine the direction of bending is reduced as the uncertainty on the sagitta
of the track is comparable to the sagitta itself. Therefore, in this energy 
range, electron events are selected without the rigidity-related selection 
criteria of Table~\ref{all_cuts}, and electrons are not distinguished from
positrons.

\placetable{elec_numb}

As a test to verify the degree of potential residual contamination of the
positron counts by interacting hadrons, the electron selection criteria
were varied from very loose to very restrictive. In all but the loosest
sets of selection criteria, the electron and positron counts were reduced
or augmented in the same proportion, indicative of little, if any, residual
contamination.

\section{Absolute Energy Spectra}

The absolute differential energy spectra of cosmic-ray primary electrons or 
positrons are obtained from the raw electron or positron counts $\Delta N$ of 
Table~\ref{elec_numb} by calculating:
\begin{equation}
j_{pri}(\bar E) \approx \frac{f_{ToA} \Delta N}
{\Delta E\:\mbox{\large $\epsilon$}\,\Omega\,A\,\Delta t} -  
j_{sec}(\bar E)
\end{equation}
with \[ \bar E = \frac{\int_{E_i}^{E_j} E\,E^{-\alpha}\,dE}
{\int_{E_i}^{E_j} E^{-\alpha}\,dE} \]
\[ \Delta E = \int_{E_i}^{E_j}
E^{-\alpha}\,dE/\bar E^{-\alpha}, \]
where an atmospheric secondary ({\it sec}) component is subtracted to obtain the
primary ({\it pri}) component. In equation~(1), $\epsilon \Omega A$ is the
instrumental acceptance, $\Delta t$ is the live time, 
$E_i$ and $E_j$ are the lower and upper bounds of each ToA energy 
interval, respectively, $\bar E$ is the weighted average ToA energy in 
this interval, $\Delta E$ is the weighted ToA energy interval, 
$f_{ToA}$ is a correction factor related to the transformation of
the energy scale to the top of the atmosphere, which will be described in
Section~\ref{atmoscor},
and $\alpha$ ($\approx 3.1$ for primary particles and 
$\approx 3.2$ for atmospheric secondaries) is the power-law spectral index.

The live time $\Delta t$ of the detector is measured in flight with a scaler
which counts clock cycles only while the instrument is available for triggers
and not busy
processing an event following a trigger. Periods of instability of the
instrument's performance are discarded, and a total live time
of $(18.04 \pm 0.04)$~hours at float altitude is obtained.
This live time includes losses due to transmission errors and data-unpacking 
errors from glitches in the electronics of individual subsystems.

\subsection{Monte Carlo Simulations\label{monte}}

Some of the quantities in equation~(1) are determined with a full Monte Carlo 
(MC) simulation of the HEAT instrument, based on the GEANT 
software package, and including the actual detector configuration and the
experimentally-determined fluctuations in detector response.
For instance, the MC simulation is used to calculate the
instrumental acceptance for elec\-trons, i.e., the pro\-duct of the absolute 
efficiency {\large $\epsilon$} and the geometrical aperture $\Omega A$.
The MC-calculated efficiencies are compared to experimental quantities
where possible, and in some cases renormalized to ensure agreement.
For example, the TRD electron selection efficiency is determined by
obtaining a set of electron events based on EC, TOF and DTH information,
and measuring the fraction of such events that satisfy the TRD likelihood 
ratio requirement. 
In addition, a
visual inspection of several hundred raw experimental and simulated electron 
events revealed that about 65\% of experimental events are accepted
by the analysis, whereas 81\% of simulated events are accepted. The
instrumental acceptance determined by the MC is therefore corrected by a
``scanning'' efficiency factor of 
{\large $\epsilon_{\mbox{\scriptsize \it scan}}$}
$= 0.65/0.81 = 0.80 \pm 0.08$, the uncertainties being
an estimate of the spread of values due to independent scannings.

Table~\ref{acc_liv} lists, for each energy interval, the acceptances 
calculated with the MC, including the
{\large $\epsilon_{\mbox{\scriptsize \it scan}}$} correction. 
The geometrical aperture, uncorrected for efficiencies, is about
495~$\mbox{cm}^2\,\mbox{sr}$. This indicates an average electron
acceptance efficiency of about 37\%. 
In Table~\ref{acc_liv}, the acceptance uncertainties are obtained by adding in
quadrature the uncertainties in the individual efficiency factors.
The increased acceptance of the highest energy bin derives from an increased
efficiency due to a different set of selection criteria used in this energy
range, as indicated in Table~\ref{all_cuts}.

\placetable{acc_liv}

A comparison of experimental and simulated $E/p$ distributions revealed
a systematic energy calibration offset in the flight data. Specifically, 
the experimentally-determined \hss

\noindent $\left| E/p \right|$ peak for electrons is 
shifted upwards by about 14\%, whereas the Monte Carlo simulation
predicts a shift of only 4\% due to the brems\-strah\-lung energy losses. The 
10\%
discrepancy appears to be due to a systematic bias in the conversion of 
the ADC values of the EC into numbers of minimum-ionizing particles. This 
energy-scale shift is corrected for in Figure~\ref{fig12}.

The measured energy spectrum is shifted without a change in spectral index
due to the finite, but roughly constant, energy resolution of the EC. 
The Monte Carlo simulation shows that of these events which have a true
initial energy in a given energy bin, a fraction of 2 to 6\% are reconstructed
by the analysis into the next higher-energy bin, while 22 to 32\% are
reconstructed into the adjacent lower-energy bin. This ``spillover'' effect
reduces the reconstructed energy spectrum by about 10\% in overall intensity as
compared to the true primary spectrum. The final results have been corrected
for this effect. Only events with ToA energies above 5~GeV, well above the 
geomagnetic rigidity cutoff at $\sim4.5$~GV, are retained in the analysis.

\subsection{Atmospheric Corrections\label{atmoscor}}

The electron spectrum measured at balloon altitudes is comprised of primary 
cosmic-ray electrons and atmospheric secondary electrons. The secondary component
arises as a result of interactions of hadrons or primary electrons within the 
atmosphere, as well as reentrant albedo electrons. Reentrant albedo particles
are significant only below the geomagnetic cutoff energy.
To determine the secondary flux generated in the atmosphere, a
Monte Carlo simulation utilizing the CERN libraries' FLUKA hadronic 
interaction algorithm (\cite{fas:flu}) is used.
For this calculation, a primary proton spectrum with a power-law index of 
2.74 is assumed (\cite{seo:prot}). Secondaries produced by primary electrons
do not contribute significantly. To account for the contribution of 
heavy primary nuclei, the simulated intensity of atmospheric secondaries is 
multiplied by 1.2 (\cite{buf:sec}). The total intensity of 10~GeV secondary 
electrons and positrons at 6~g/cm$^2$ is found to be 
0.006~(GeV s sr m$^{2}$)$^{-1}$. This corresponds to overall corrections to 
the positron intensity of 20--30\%, and to the electron intensity of 1--2\%, 
at the energies of interest. The 
absolute atmospheric intensities at an average atmospheric depth of 
5.7~g/cm$^2$ for electrons and positrons are also presented in 
Table~\ref{atm_sec}, for each energy interval.

\placetable{atm_sec}

\begin{figure}
\vspace{-0.5cm}
\hbox to \textwidth{\hss
\epsfxsize=0.4\textwidth
\epsfbox[180 280 430 570]{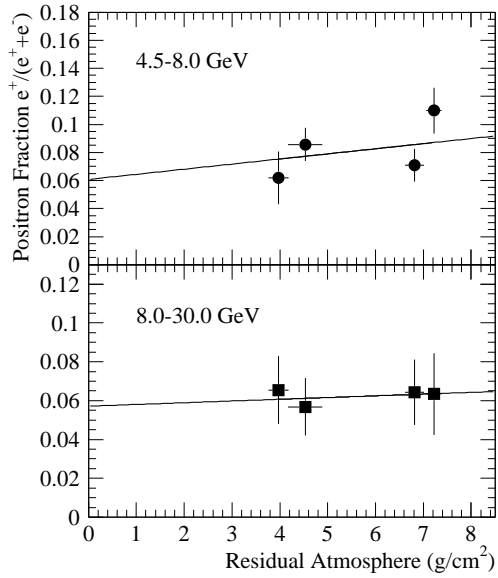}
\hss}
\caption{\it Positron fraction as a function of atmospheric depth
\label{fig13} }
\end{figure}

As a test of the reliability of the MC calculation, atmospheric growth curves 
for electrons and positrons are measured during the flight. The positron 
fraction as a function of depth in the atmosphere is shown in 
Figure~\ref{fig13}, for two separate energy intervals.
The flight data are divided into four altitude intervals of 
4.0, 4.5, 6.8, and 7.2~g/cm$^{2}$ average depth, respectively. 
The error bars represent statistical 
fluctuations, dominated by the number of positrons in each energy interval.
The positron fraction at the top of the atmosphere is 
obtained by linear extrapolation of the experimental growth curves.  
From the data on Figure~\ref{fig13}, one can determine the
secondary-to-primary ratio $(e^++e^-)_{sec}/(e^++e^-)_{pri}$ as a function of
atmospheric depth. In Table~\ref{sec_prim_ratio}, this ratio at a depth of 
6~g/cm$^2$ is given and compared to the corresponding results of the MC
calculations. The uncertainties indicated are 
statistical. The MC simulation is consistent 
with the measured secondary-to-primary ratios within errors.

\placefigure{fig13}

\placetable{sec_prim_ratio}

The energy of a primary electron detected at a residual atmospheric depth of
$t$ radiation lengths is reduced from the energy at the top of the atmosphere
due to bremsstrahlung energy losses. If the primary spectrum has the form of a
power law $E^{-\alpha}$, the detected energy spectrum will, for a given depth
$t$, retain the same power law but will be reduced in intensity by a factor
$\alpha^{-t/\ln2}$. This factor is derived by taking the statistical
fluctuations in the radiation loss mechanism into account 
(\cite{ros:cr}, \cite{sch:mid}). For the HEAT flight described here, the
atmospheric depth varies considerably throughout the flight. Therefore, we
correct the energy for each detected electron or positron by a factor 
$f = \alpha^{t/\alpha \ln2}$, where $t$ is the current
atmospheric depth, and $\alpha = 3.1$ is taken for the spectral index.
Typically, this amounts to a $\sim$~5-10\% shift in the energy scale. 
Because equation~(1) is used to calculate a {\it differential} intensity,
the energy scale shift necessitates a further correction factor 
$f_{ToA} = \langle f \rangle = \langle {dE_{ToA} \over dE} \rangle
= (1.086 \pm 0.020)$, where $f$ is averaged over the flight.

\section{Results and Discussion}

\begin{figure}
\vspace{-0.5cm}
\hbox to \textwidth{\hss
\epsfxsize=0.4\textwidth
\epsfbox[180 280 430 560]{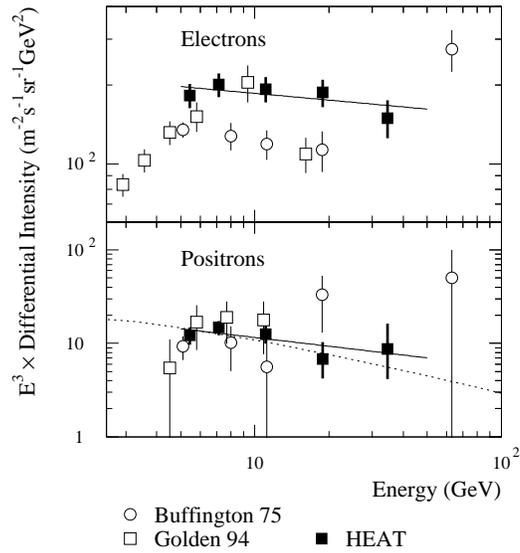}
\hss}
\caption{\it Absolute differential energy spectra for electrons and
positrons; the dotted line is a prediction from Strong et al. 1997,
and the solid lines are power-law fits to the HEAT results
\label{fig14} }
\end{figure}

\begin{figure}
\vspace{-0.5cm}
\hbox to \textwidth{\hss
\epsfxsize=0.8\textwidth
\epsfbox[50 290 550 580]{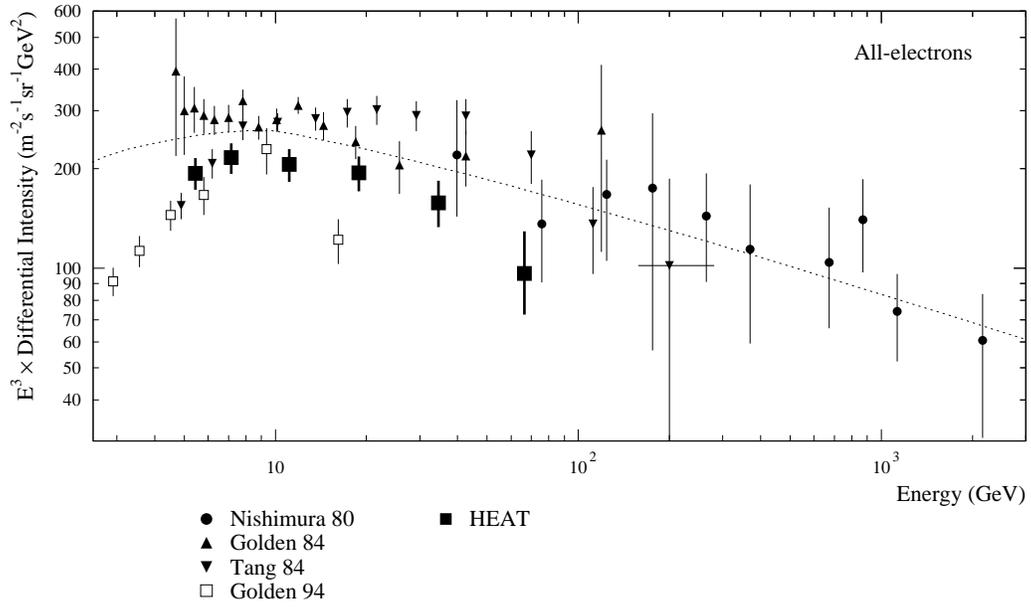}
\hss}
\caption{\it Absolute differential energy spectrum for all electrons 
$({\rm e^+ + e^-})$; the dotted line is a parametrization from 
Strong et al. 1997
\label{fig15} }
\end{figure}

The absolute differential intensities of electrons and positrons are obtained 
from equation~(1) 
using Table~\ref{elec_numb}, which lists the number of particles counted in each
energy interval $\Delta N$, Table~\ref{acc_liv} which tabulates the acceptances 
{\large $\epsilon$} $\!\Omega A$, live time $\Delta t$ and average ToA
energy correction factor $f_{ToA}$, as well as
$\bar E$ and $\Delta E$ (computed with $\alpha = 3.1$), and
Table~\ref{atm_sec} which gives the atmospheric contribution. All of the 
statistical uncertainties associated with these quantities are also contained 
in these tables. The electron, positron, and all-electron differential
intensities are
listed in Table~\ref{abs_flux}. The uncertainties are statistical, computed 
by adding in quadrature all of the contributing uncertainties.
The results are plotted in Figures~\ref{fig14} and \ref{fig15},
scaled by $E^3$ for clarity, together with other measurements of electron 
and positron energy spectra
(\cite{buf:pos,nis:ele,gol:ele,tan:the,go2:pos}).

\placetable{abs_flux}

\placefigure{fig14}

\placefigure{fig15}

Besides the uncertainty in determining the instrumental efficiencies
discussed in Section \ref{monte}, there are two additional sources of 
systematic errors. First, the necessity of
applying an energy-shift correction to ensure agreement with the momentum 
indicates a systematic uncertainty in the absolute energy calibration of
a few percent, which translates into a $\sim$10\% uncertainty in the overall 
intensity normalization due to the $\alpha = 3.1$ power law of the spectrum. 
Second, uncertainties in the normalization of the proton spectrum
translate into a $\sim 25\%$ systematic uncertainty in simulated atmospheric 
secondary intensities; this has a 0.5\% effect on the electron and all-electron 
intensities, but results in 10\% shifts in the positron intensity normalization.
Thus, the estimated systematic uncertainty from these sources is 
10\% for the electron and all-electron intensities, and 14\% for the positron 
intensity.

In Figure~\ref{fig14}, we compare our results with previous measurements for 
which absolute intensities have been given for positrons and electrons
separately. We find that the energy spectrum of positrons reported 
by \cite{buf:pos} and \cite{go2:pos} is generally consistent with our data, 
although our results show much improved statistical error limits and permit
a better estimate of the slope of the energy spectrum.
The intensity of positrons of interstellar secondary origin can be
calculated, provided the spectrum of parent protons is known, and the
parameters of galactic propagation are specified. Such a calculation has been
done by \cite{pro:pos}, and, more recently, by \cite{str:pos}. We show
in Figure~\ref{fig14} the results of Strong et al. 1997 as a dotted line, and
notice the good agreement with our data.
In Figure~\ref{fig15},
we compare the all-electron energy spectrum ${\rm (e^+ + e^-)}$ derived from 
our data with previous results, also including measurements which did not
employ a magnet spectrometer for charge separation.
We arbitrarily chose to include only results published since 1980.
The dotted line represents a parametrization of the all-electron spectrum used
by Strong et al. 1997.
We find that all results exhibit similar spectral slopes over the energy range 
of concern, but that our overall intensity is lower than the average of the 
previous investigations by about 35\%. There are two possible
contributors to this systematic discrepancy. First, the absolute energy
calibration for the experiments is likely to be uncertain by about 10\%,
leading to a possible systematic uncertainty in the intensity of the order of 
20\%. Second,
the assessment of the absolute detection efficiency of the instrument is
notoriously difficult for all such detectors. Its determination does
involve some intuitive judgment, although the more recent investigations,
such as ours, benefit in this respect from more sophisticated Monte Carlo
simulations than were previously available.

\placetable{pow_fit}

The individual spectral indices for
po\-si\-trons and electrons are $\alpha = 3.3 \pm 0.2$ and $3.09 \pm 0.08$,
respectively, over the energy range 5.0 to 50~GeV, 
as indicated in Table~\ref{pow_fit} and represented by solid
lines in Figure~\ref{fig14}. Thus, the positron
spectrum appears to be slightly steeper than that of electrons. If all
positrons are of interstellar secondary origin, one should expect that
radiative energy losses eventually lead to a power-law index that is larger
by unity than that of the production spectrum, which follows the ambient
spectrum of the parent nuclei, i.e., $\alpha \approx 3.7$ for nuclear spectra
of the form $E^{-2.7}$. The spectrum of electrons, on the other hand, should
steepen less strongly, to a final slope with $\alpha \approx 3.1$ if electrons 
are mostly produced at the same primary sources as nuclei, and with the same
source spectrum characterized by $\alpha \approx 2.1$. However, these final 
spectral
slopes may not be reached in the energy region below $\sim 100$~GeV. Thus, it
would seem proper to fit our data not to a single power law but rather to a
spectral form that reflects a transition from the source spectrum (modified
by solar modulation) to a
spectrum that is fully steepened due to radiative energy losses. Such an
analysis had previously been performed on the all-electron spectrum by 
\cite{pri:the} and \cite{tan:the}. Here we will not 
entertain this analysis for our data, as we soon should be able to improve the
statistical quality of the results through the inclusion of data from an
additional balloon flight. Qualitatively, we may just conclude that the
slightly steeper spectrum of positrons, as compared to that of electrons,
is to be expected if positrons are predominantly secondary particles.
This conclusion must be confirmed by additional data and by an
extension of the measurement over a wider energy range.

As mentioned above, a comparison with the calculations of \cite{pro:pos} and 
Strong et al. 1997 indicates also that
the absolute intensity of positrons is close to what may be expected
if indeed all positrons are generated subsequent to nuclear interactions in
the interstellar medium. 
More detail on the relative
origin of electrons and positrons, and limits on the possible primary
contribution to the positron flux, can be obtained if the relative
intensity of positrons, i.e., the positron fraction
${\rm e^+/(e^+ + e^-)}$ (which does
not depend on the absolute efficiency of the detector), is investigated as
a function of energy.  We have presented the results of such an investigation
in previous papers (\cite{bar:prl,bar:apjl}).

The results described in this paper have been obtained with the first
balloon flight of the HEAT spectrometer. We believe that the high level of
hadronic background rejection achieved with this instrument has led to data
of high quality, limited only by the statistical uncertainties of a single
balloon flight. A second balloon flight which has been carried out
successfully from Lynn Lake, Manitoba in 1995 August, covers the energy
range from $\sim 1$~GeV upwards. The analysis of the combined data set is
presently in progress, and results will soon be reported.

\acknowledgments

We gratefully acknowledge assistance from D.~Bonasera, E.~Drag, 
W.~Johnson, P.~Koehn, D.~Kouba, R.~Northrop, and J.~Robbins.  
We are grateful to M.~Israel for helpful discussions on atmospheric secondary
production.
We also thank the NSBF crew that has supported the balloon flight.  
This work was supported by NASA grants 
NAGW-\-1035, NAGW-\-1995, NAGW-\-2000, NAGW-\-4737, 
NAG5-\-5059, \hss

\noindent NAG5-\-5069, and NAG5-\-5070,
and by financial assistance 
from our universities. D.~E. and E.~T. acknowledge support from the \hss

\noindent NASA graduate student researchers program.

% Insert references here:

% Insert tables here:
% Table of final cuts (cleanliness, electron, interacting proton):
\clearpage
\begin{deluxetable}{l}
\tablecaption{Data selection criteria \label{all_cuts}}
\tablewidth{0pt}
\startdata
\vspace{-2em}
\\
\hline
\hline
{\bf Cleanliness requirements} \\
DTH filter algorithm passed \\
DTH rigidity reconstruction algorithm passed \\
$N\!B \ge 9$ (Number of tubes in DTH bending view used in fit) \\
$N\!N\!B \ge 4$ (Number of tubes in DTH non-bending view used in fit) \\
$\chi^2 < 10$  (DTH rigidity goodness of fit) \\
$D\!E\!V\!\_X < 0.080$~cm (Average residual in $X$ (non-bending) projection) \\
$D\!E\!V\!\_Y < 0.014$~cm (Average residual in $Y$ (bending) projection)  \\
$D\!E\!V\!\_Z < 0.020$~cm (Average residual in $Z$ (vertical) projection) \\
$\int\!B\,dl > 2.2$~kG~m (Integrated B-field over the track 
length)\tablenotemark{a} \\
$M\!\!D\!R > 60$~GV\tablenotemark{a} \\
$M\!\!D\!R/ \left| R \right| > 4$\tablenotemark{a} \\
$\left|E/p\right| < 3.0$\tablenotemark{a} \\
$0.70 < \beta < 1.65$ (Velocity range, downward-going) \\
$\left|T\!O\!F\!\_X - D\!T\!\_X\right| < 20$~cm (Agreement between DTH track
and TOF timing) \\
Propagated DTH track traverses both TOF and EC \\
Propagated DTH track traverses $\ge 4$ TRD chambers \\
\hline
{\bf Electron selection criteria} \\
$0.68 < Z < 1.45$ (Charge) \\
EC starting depth $t < 0.89$ radiation lengths \\
$\chi^2 < 2.6$ (EC goodness of fit) \\
$\log_{10}\left[(L_{ep}^{PHA} - \sigma_{L_{ep}^{PHA}}) \cdot
(L_{ep}^{TS} - \sigma_{L_{ep}^{TS}})\right] > 3$ (TRD likelihood ratio)\\
\hline
{\bf Interacting proton selection criteria} \\
Same as electron selection criteria except: \\
$\log_{10}\left[(L_{ep}^{PHA} + \sigma_{L_{ep}^{PHA}}) \cdot
(L_{ep}^{TS} + \sigma_{L_{ep}^{TS}})\right] < -1$ (TRD likelihood ratio)\\
\enddata
\tablenotetext{a}{This selection is not used in the 50--100~GeV energy range.}
\end{deluxetable}

%Table of raw numbers of electrons, from template fits:
\clearpage
\begin{deluxetable}{cccc}
\tablecaption{Raw numbers of electrons \label{elec_numb}}
\tablewidth{0pt}
\tablehead{
\colhead{$E_{ToA}$ (GeV)} &
  \colhead{\# Electrons $\Delta N_{\rm e^-}$} &
    \colhead{\# Positrons $\Delta N_{\rm e^+}$} &
      \colhead{\# All-electrons $\Delta N_{\rm e}$}}
\startdata
5.0 -- 6.0 &
  $1231 \pm 36$ &
    $107 \pm 11$ &
      $1338 \pm 37$ \\
6.0 -- 8.9 &
  $1781 \pm 43$ &
    $161 \pm 14$ &
      $1942 \pm 45$ \\
8.9 -- 14.8 &
  $918 \pm 31$ &
    $75^{+10}_{-9}$ &
      $993 \pm 33$ \\
14.8 -- 26.5 &
  $340 \pm 20$ &
    $18.5^{+5.8}_{-3.9}$ &
      $359 \pm 20$ \\
26.5 -- 50.0 &
  $75^{+10}_{-9}$ &
    $6.1^{+3.7}_{-2.1}$ &
      $81^{+10}_{-9}$ \\
50.0 -- 100.0 & & &
      $19.2^{+5.7}_{-4.0}$ \\
\enddata
\end{deluxetable}

% Table of Ebar, DeltaE, live time, and acceptances:
\clearpage
\begin{deluxetable}{ccccc}
\tablecaption{Energy intervals and effective acceptances
\label{acc_liv}}
\tablecomments{the live time is $\Delta t = (18.043 \pm 0.036)$~hr, and the
average ToA energy correction factor is $f_{ToA}=(1.086\pm 0.020)$}
\tablewidth{0pt}
\tablehead{
\colhead{$E_{ToA}$ (GeV)} &
  \colhead{$\bar E$ (GeV)} &
    \colhead{$\Delta E$ (GeV)} &
      \colhead{{\large $\epsilon$} $\!\Omega A\:(\mbox{cm}^2\,\mbox{sr})$}}
\startdata
5.0 -- 6.0 &
  5.45 &
    0.991 &
      $181 \pm 18$ \\
6.0 -- 8.9 &
  7.16 &
    2.78 &
      $193 \pm 19$ \\
8.9 -- 14.8 &
  11.1 &
    5.50 &
      $194 \pm 20$ \\
14.8 -- 26.5 &
  18.9 &
    10.7 &
      $189 \pm 19$ \\
26.5 -- 50.0 &
  34.5 &
    21.1 &
      $160 \pm 16$ \\
50.0 -- 100.0 &
  66.4 &
    44.0 &
      $207 \pm 21$ \\
\enddata
\end{deluxetable}

% Table of atmospheric secondary fluxes:
\clearpage
\begin{deluxetable}{ccc}
\tablecaption{Atmospheric secondary electron and positron intensities at
5.7~{\rm g/cm}$^2$
\label{atm_sec}}
\tablewidth{0pt}
\tablehead{
\colhead{$E_{ToA}$ (GeV)} &
  \colhead{$j_{sec}^{-}
  \!({\mbox{m}}^2\,\mbox{s}\,\mbox{sr}\,\mbox{GeV})^{-1}$} &
    \colhead{$j_{sec}^{+}
    \!({\mbox{m}}^2\,\mbox{s}\,\mbox{sr}\,\mbox{GeV})^{-1}$}}
    \startdata
5.0 -- 6.0   & $(1.8 \pm 0.6)\times 10^{-2}$ & $(2.4 \pm 0.6)\times 10^{-2}$ \\
6.0 -- 8.9   & $(7.6 \pm 2.3)\times 10^{-3}$ & $(9.7 \pm 2.2)\times 10^{-3}$ \\
8.9 -- 14.8  & $(2.1 \pm 0.6)\times 10^{-3}$ & $(2.5 \pm 0.7)\times 10^{-3}$ \\
14.8 -- 26.5 & $(4 \pm 1)\times 10^{-4}$ & $(5.3 \pm 1.1)\times 10^{-4}$ \\
26.5 -- 50.0 & $(7 \pm 2)\times 10^{-5}$ & $(9 \pm 2)\times 10^{-5}$ \\
50.0 -- 100.0 & $(1.1 \pm 0.7)\times 10^{-5}$ & $(1.2 \pm 0.7)\times 10^{-5}$ \\
\enddata
\end{deluxetable}

% Table of secondary-to-primary ratios:
\clearpage
\begin{deluxetable}{ccc}
\tablecaption{Secondary-to-primary electron ratios at 6~{\rm g/cm}$^{2}$ 
\label{sec_prim_ratio}}
\tablewidth{0pt}
\tablehead{
\colhead{Energy (GeV)} &
  \colhead{MC sec./pri.} &
    \colhead{HEAT sec./pri.}}
\startdata
4.5 -- 6.0   & 0.04 $\pm$ 0.01   & 0.04 $\pm$ 0.03  \\ 
6.0 -- 8.9   & 0.04 $\pm$ 0.01   & 0.04 $\pm$ 0.03  \\ 
8.9 -- 14.8  & 0.04 $\pm$ 0.01   & 0.01 $\pm$ 0.03  \\ 
14.8 -- 25.6 & 0.05 $\pm$ 0.01   & 0.01 $\pm$ 0.03  \\ 
25.6 -- 50.0 & 0.06 $\pm$ 0.02   &         \\ 
\enddata
\end{deluxetable}

% Table of Electron Fluxes:
\clearpage
\begin{deluxetable}{clll}
\tablecaption{Differential intensities of electrons, in
$({\mbox{\rm m}}^2\,\mbox{\rm s}\,\mbox{\rm sr}\,\mbox{\rm GeV})^{-1}$
\label{abs_flux}}
\tablewidth{0pt}
\tablehead{
\colhead{$\bar E$ (GeV)} &
  \colhead{$j_{pri}^{-}(\bar E)$} &
    \colhead{$j_{pri}^{+}(\bar E)$} &
      \colhead{$j_{pri}^{\pm}(\bar E)$}}
\startdata
5.45 &
  $1.13 \pm 0.12$ &
    $0.076 \pm 0.016$ &
      $1.20 \pm 0.13$ \\
7.16 &
  $0.548 \pm 0.057$ &
    $0.0405 \pm 0.0070$ &
      $0.589 \pm 0.062$ \\
11.1 &
  $0.141 \pm 0.016$ &
    $(9.2^{+2.1}_{-2.0}) \times 10^{-3}$ &
      $0.151 \pm 0.017$ \\
18.9 &
  $0.0278 \pm 0.0033$ &
    $(1.00^{+0.52}_{-0.38}) \times 10^{-3}$ &
      $0.0288 \pm 0.0035$ \\
34.5 &
  $(3.64^{+0.62}_{-0.58}) \times 10^{-3}$ &
    $(2.1^{+1.9}_{-1.1}) \times 10^{-4}$ &
      $(3.84^{+0.64}_{-0.60}) \times 10^{-3}$ \\
66.4 &    &    &
      $(3.30^{+1.11}_{-0.83}) \times 10^{-4}$ \\
\enddata
\end{deluxetable}

% Table of power-law fit parameters:
\clearpage
\begin{deluxetable}{ccc}
\tablecaption{Power-law fits to the differential spectra
\label{pow_fit}}
\tablewidth{0pt}
\tablehead{
\colhead{$J_\circ E^{-\alpha}$} &
  \colhead{$\alpha$} &
    \colhead{
    $J_\circ\:(\mbox{GeV}^{-1}\,\mbox{m}^{-2}\,\mbox{s}^{-1}\,\mbox{sr})$}}
\startdata
Electrons &
  $3.086 \pm 0.081$ &
    $227 \pm 45$ \\
Positrons &
  $3.31 \pm 0.23$ &
    $24 \pm 12$ \\
\enddata
\end{deluxetable}

% Insert figure captions here:
\clearpage

\onecolumn

\clearpage

\clearpage

\clearpage

\clearpage

\clearpage

\clearpage

\clearpage

\end{document}